\documentclass[twocolumn,amsmath,amssymb]{revtex4}

\usepackage{axodraw}

\newcommand{\del}{\partial}
\newcommand{\Mp}{M_{\rm P}}
\newcommand{\Ms}{M_{\rm s}}
\newcommand{\gsim}{\lower.7ex\hbox{$\;\stackrel{\textstyle>}{\sim}\;$}}
\newcommand{\lsim}{\lower.7ex\hbox{$\;\stackrel{\textstyle<}{\sim}\;$}}

\begin{document}

\preprint{IPPP/0408}
\preprint{DCPT/0416}

\title{ Naturally small Dirac neutrino masses in supergravity}

\author{Steven Abel$^{\, a}$, Athanasios Dedes$^{\, a}$ and  
Kyriakos Tamvakis$^{\, b}$}

%\affiliation{$^a$ Center of Particle Theory, 
%University of Durham, Durham DH1 3LE, UK }

\affiliation{$^a$Institute for Particle Physics Phenomenology (IPPP), 
Durham DH1 3LE, UK }

\affiliation{$^b$Physics Department, University of Ioannina, 
GR 451 10, Ioannina, Greece  }

\begin{abstract}
\noindent We show that Dirac neutrino masses of the right size
%automatically 
can arise from the K\"ahler potential of supergravity.
 They are proportional to
the supersymmetry and the electroweak breaking scales.  We find that
they have the experimentally observed value provided that the
ultraviolet cut-off of the Minimal Supersymmetric Standard Model 
(MSSM) is between the Grand Unification
(GUT) scale and the heterotic string scale. If lepton number is not
conserved, then relatively suppressed  Majorana
masses can also be present, resulting in  pseudo-Dirac neutrino masses. 
%with generally small splitting.
%, resulting in pseudo-Dirac neutrino masses.
%, with a Majorana 
%component of $10^{-4}$ eV or less.
\end{abstract}

\pacs{04.65.+e, 12.15.Ff, 12.60.Jv, 14.60.Pq, 14.60.St}

\maketitle

\section{Introduction}
%\label{sec:intro}

We have recently learned a
great deal about mixing in the neutrino sector \cite{nureview}. 
However we have thus
far learned relatively little about why the neutrino masses are so small, or
their relation to the other much higher scales in particle physics.
The presence of such vastly different mass scales remains a great
puzzle. 
The current favoured explanation for small neutrino masses is the
{}``see-saw'' mechanism \cite{seesaw}. 
In this picture a large Majorana mass for
the right handed neutrino suppresses the mass of the light states, 
and the active neutrinos we observe today are therefore almost
pure Majorana. 
%A possibly unappealing feature of the see-saw mechanism 
%is that, by introducing
%a high Majorana scale, we are merely trading a small unexplained mass
%scale for a very large one (although the latter is perhaps closer to 
%scales we might be thinking about for other problems). 
However, in this 
framework there is
no room for Dirac or pseudo-Dirac neutrinos, and so it is worth
examining alternative ways to generate neutrino masses.

There is one other instance in supersymmetry where it was possible
drastically to suppress a mass scale, the solution of the
$\mu$- problem by Giudice and Masiero \cite{GM}. The $\mu H_{u}H_{d}$ term
in the superpotential is a mass term for the Higgs fields required
for electroweak symmetry breaking. The parameter $\mu$, which has dimensions of
mass, therefore has to be of the order of $1$ TeV. But in global supersymmetry 
it is apparently independent of the 
supersymmetry breaking terms which also have to
be of order $1$ TeV, appearing as it does in the superpotential.
At first glance there is no connection between supersymmetry
breaking and the parameter $\mu$. However the problem is resolved if the
$H_{u}H_{d}$ interaction appears in the K\"ahler potential of supergravity
rather than the superpotential. Then an effective $\mu-$term is generated
only upon supersymmetry breaking and is of the order of the gravitino mass
$m_{3/2}\sim1$ TeV. The crucial ingredient of this solution to the 
$\mu $ problem is the absence of this term in the superpotential
of unbroken supergravity, 
and its subsequent generation through an analogous coupling in the 
K\"ahler potential, once supersymmetry is broken.

Could such a K\"ahler suppression be responsible for the smallness of
neutrino masses as well? The numbers certainly suggest that it could
be as has been occasionally noted in the literature in the context of 
global supersymmetry or 
globally supersymmetric approximations
to supergravity~\cite{Hall,Borzumati,Navarro,Kitano,Arnowitt}.
Consider for example a contribution to the K\"ahler potential of the
form \[
K\,\supset\,\frac{LH_{u}\bar{N}}{M}+\frac{LH_{d}^{*}\bar{N}}{M}+{\rm
H.c.}\] where $\bar{N}$ is the right handed neutrino and $M$ is the scale at
which higher dimensional operators first make their appearance in the
K\"ahler potential. For the sake of argument assume that $M =
\Mp=(8\pi G_N)^{-1/2}= 2.44\times 10^{18}$ GeV. One would expect the
effective neutrino mass to be suppressed by a factor $ m_{3/2}/M$
which (taking $\langle H_{u}\rangle=m_{\rm top}$) gives a neutrino
mass $ m_{\nu}\sim10^{-4}\,{\rm eV}.$ This is rather small but
intriguingly quite close to the measured~\footnote{We will throughout
be assuming that the measured mass-squared differences are indicative
of the actual masses. We focus on the atmospheric neutrino mass.}  
value of $(0.04-0.05)~{\rm eV}$ (within
1$\sigma$). Even more intriguingly, the measured value corresponds to
taking $M=5\times10^{15}$ GeV, just below the GUT scale. We think that
this coincidence deserves more careful inspection in the context of full
supergravity~\cite{Nilles}.

The above operators are expected to be generated in various ways (perhaps 
from some kind of GUT theory or by the underlying string theory) 
and so $M$ does not have to be close 
to $\Mp$. 
%This is especially true nowadays, when we know that various 
%kinds of string set-ups allow the string scale to be 
%much lower than $\Mp$ provided there are large compact dimensions. 
Because of this the scale $M$ (when it was not set by some 
model building assumption or other) has always
been treated as a moveable parameter. In this letter 
we take a more phenomenological approach. If the operators 
above are indeed responsible for the neutrino masses, what does the 
scale $M$ of new physics have to be? 
%This is a question that can 
%quite accurately be 
%answered in supergravity and we will find an answer  
%that differs by two orders of magnitude from the naive 
%expectation above. 
{Exploiting supergravity as a possible breakdown scenario of
supersymmetry, we find that within this scenario, the scale $M$ may
differ by two orders of magnitude from the naive expectation above.}
Indeed, for gravitino masses of $100{\rm ~GeV}<m_{3/2}<10{\rm ~TeV}$,
the correct mass automatically arises from the general couplings of
supergravity if the scale $M$ is in the range
%%%%%%%%%%%%%%%%%%%%%
\begin{eqnarray}
M = (4\times 10^{16}-5\times 10^{17}){\rm \, GeV.}
\nonumber 
\end{eqnarray}
%%%%%%%%%%%%%%%%%%%%%
This range, remarkably, is between the GUT scale and 
the heterotic string scale of old. 
At tree level the relation for the latter 
is $\Ms=g_{GUT} \Mp \approx 10^{18}$ GeV
if $\alpha_{GUT}=1/24$. Including threshold effects in the 
${\overline{MS}}$ scheme gives \cite{kaplunovsky}, 
$\Ms = 3.8\times10^{17}{\rm \, GeV}$.
%
%\setcounter{equation}{0}
%\section{Fermion masses in Supergravity}
%\label{sec:sec2}
%
To find this result, we need to consider the contributions to fermion 
masses in full supergravity.

\section{Fermion Masses in Supergravity}

Consider a set of chiral superfields $\{S_i, y_\alpha\}$.
The fields $S_i$ are those fields of the hidden sector that are 
responsible for the spontaneous breaking of supergravity. 
They are assumed to be singlets of the gauge group, and we make no other 
assumptions about them or their superpotential, apart from that they 
eventually acquire a v.e.v. of order $S_i\simeq M$.
It is convenient to set $S_i = M \sigma_i$.
The superfields  $y_\alpha$ are those of the observable
sector, namely  $y_\alpha = \{Q,\bar{U},\bar{D},L,\bar{E},\bar{N},
H_u,H_d\}$. 
%Demanding stability of the hierarchy between the high
%scale $M$ and the electroweak scale $M_W$, the 
% I REMOVED THE ABOVE - THE BELOW IS ALREADY THE MOST GENERAL?
The most general
superpotential, $W$, and K\"ahler potential, $K$, read~\cite{SW}
%%%%%%%%%%%%%%%%%%%%%%%%%%%%%%%
\begin{eqnarray}
W(\sigma,y) \ &=&  W^{\rm (h)}(\sigma) 
+ W^{\rm (o)}(\sigma,y) \;, \label{eq1} \\[2mm]
K(\sigma,\sigma^*,y,y^\dagger) \ &=& \  K^{\rm (h)}(\sigma,\sigma^*)
 + K^{\rm (o)}(\sigma,\sigma^*,y,y^\dagger)\;,
\label{eq2}
\end{eqnarray}
%%%%%%%%%%%%%%%%%%%%%%%%%%%%%%%5
where the superscript ${\rm (h)}$ and ${\rm (o)}$ denote hidden or 
observable superpotential and K\"ahler potentials, respectively.
If local supersymmetry is spontaneously broken  
then the visible matter fermions  have a Lagrangian 
of the form ~\cite{WB,sugra}~\footnote{Here $\chi_\alpha$ are the fermion
superpartners of the scalar fields $y_\alpha$ in
the observable sector.} 
%%%%%%%%%%%%%%%%%%%%%%%%%%%%%%
\begin{eqnarray}
{\cal L}
\ =\ i \,  g_{\alpha \beta^*} \bar{\chi}_\beta  \bar{\sigma}^\mu \partial_\mu  
\chi_\alpha 
\  - \  \biggl ( m_{\alpha\beta} \chi^\alpha \chi^\beta + {\rm H.c}
 \biggr ),
\label{fmatter}
\end{eqnarray}
%%%%%%%%%%%%%%%%%%%%%%%%%%%%%%%%%%%%%
where
%%%%%%%%%%%%%%%%%%%%%%%%%%%%%%
%\begin{eqnarray}
$g_{\alpha \beta^*} \ =\ \frac{\del^2 K}{\del y^\alpha \del y^{\beta*}} 
\ \stackrel{\rm Eq.(\ref{eq2})}{=} \
 \frac{\del^2 K^{\rm (o)}}{\del y^\alpha \del y^{\beta*}}\;$
%\end{eqnarray}
%%%%%%%%%%%%%%%%%%%%%%%%%%%%%%
is  the K\"ahler metric.
The fermion fields $\chi^\alpha$ in
Eq.(\ref{fmatter}) need not be in the canonical basis. 
Nevertheless
as is known from derivations of higher order operators in the 
K\"ahler potential (in 
for example string theory in Ref.\cite{Antoniadis}),
the various symmetries of 
the theory dictate that their coefficients are
of order one in the canonical basis.
{For simplicity reasons, we  confine our numerical discussions 
to that case, namely $g^{ij*}=g^{\alpha\beta*}=1$.}

%Thus without loss of generality 
%we can confine our numerical discussions 
%to that case, namely $g^{ij*}=g^{\alpha\beta*}=1$.

%Although one can obtain enhanced neutrino masses by arranging 
%the K\"ahler metric to be
%close to singular, we shall not need to make use of such a peculiar
%mechanism in what follows.
%%%%%%%%%%%%%%%%%%%%%%

With a general K\"ahler metric, fermion masses in supergravity read,
%%%%%%%%%%%%%%%%%%%%%%%%%%%%%
\begin{widetext}
%%%%%%%%%%%%%%%%%%%%%%%%%%%%%
\begin{eqnarray}
m_{\alpha\beta} \ &=& \ \frac{1}{2}\: \Biggl \{ 
\frac{\del^2 W^{\rm (o)}}{\del y^a\del y^\beta}
%+ \frac{\del^2 K^{\rm (o)}}{\del y^\alpha \del y^\beta} 
%%%%%
- g^{\gamma\delta*}\frac{\del^3 K^{\rm (o)}}{\del y^\alpha\del 
y^\beta \del y^{\delta*}}
\frac{\del W^{\rm (o)}}{\del y^\gamma}
%%%%% 
-  \frac{1}{M^2} \biggl [
g^{ij*} \frac{\del^3 K^{\rm (o)}}{\del y^\alpha \del y^\beta \del
\sigma^{j*}}\frac{\del W^{\rm (h)}}{\del \sigma^i}\biggl ]
\nonumber \\[2mm]& &   
%%%%%
\;\;\;\;\;\;\;\;\;\; -\frac{1}{M} \biggl [ g^{\gamma i*} 
\frac{\del^3 K^{\rm (o)}}{\del y^\alpha\del y^\beta \del \sigma^{i*}}
\frac{\del W^{\rm (o)}}{\del y^\gamma} +
g^{i\delta*} 
\frac{\del^3 K^{\rm (o)}}{\del y^\alpha\del y^\beta \del y^{\delta*}}
\frac{\del W^{\rm (h)}}{\del \sigma^i} \biggr ] 
%%%%%%%%%%%%%%%%%%%%%%%%%%%%%%%%%%%%%%%%%%%%%%%%%%%%%%%%%%%%%
%\nonumber \\[2mm]&-&
 \Biggr \} 
\nonumber \\[2mm]&-&
\frac{m_{3/2}}{2}\: \Biggl \{ 
%%%%%
g^{\gamma\delta*}\frac{\del^3 K^{\rm (o)}}{\del y^\alpha\del 
y^\beta \del y^{\delta*}}
\frac{\del K^{\rm (o)}}{\del y^\gamma}
%%%%%
- \frac{\del^2 K^{\rm (o)}}{\del y^\alpha \del y^\beta} 
%%%%% 
+  \frac{1}{M^2} \biggl [
g^{ij*} \frac{\del^3 K^{\rm (o)}}{\del y^\alpha \del y^\beta \del
\sigma^{j*}}\frac{\del K^{\rm (h)}}{\del \sigma^i}\biggl ]
\nonumber \\[2mm]  & & 
%%%%%
\;\;\;\;\;\;\;\;\;\; +\frac{1}{M} \biggl [ g^{\gamma i*} 
\frac{\del^3 K^{\rm (o)}}{\del y^\alpha\del y^\beta \del \sigma^{i*}}
\frac{\del K^{\rm (o)}}{\del y^\gamma} +
g^{i\delta*} 
\frac{\del^3 K^{\rm (o)}}{\del y^\alpha\del y^\beta \del y^{\delta*}}
\frac{\del K^{\rm (h)}}{\del \sigma^i} \biggr ] 
%%%%%
%\nonumber \\[2mm]&-&
 \Biggr \} \;.
\label{matf}
\end{eqnarray}
%%%%%%%%%%%%%%%%%%%%%%%%%%%%%%%%%%%%
\end{widetext}
%%%%%%%%%%%%%%%%%%%%%%%%%%%%%%%%%%%%%
In the above $m_{3/2}$ is the gravitino mass given by 
\begin{eqnarray}
m_{3/2} = \langle \frac{W^{\rm (h)}}{\Mp^2} 
\: {\rm exp}(K^{\rm (h)}/2\Mp^2) \rangle \; ,
\end{eqnarray}  
and we have taken the flat limit, $\Mp \to \infty$ and 
$m_{3/2} \to { const}$.
We should remark here that we have made no other approximations in deriving
Eq.(\ref{matf}). Contributions to the visible fermion masses in
Eq.(\ref{matf}) arise from both the hidden and the observable
sectors. We have divided the contributions to the fermion masses into
two classes:

\indent {\it i)} terms which are not proportional to the gravitino mass
and survive in the global supersymmetry limit 
$m_{3/2}\to 0$, $m_{3/2} \Mp \to const.$
(the first two lines of Eq.(\ref{matf})).
Of these terms the first can be recognized as the 
standard  term present in global supersymmetry.
The second term arises purely from the observable  sector.
It was used by the authors of  Ref.~\cite{Navarro} in order
to induce Majorana neutrino masses from dimension six
K\"ahler operators.
{In our scenario}, the third  term in the first line of  Eq.(\ref{matf}) 
is precisely the term that produces the dominant
contribution to the neutrino masses. Note that this term has
{\em not} previously been considered in the context of neutrino masses, 
and can significantly change 
%(by orders of magnitude)
any estimates that one might make within the framework of Supergravity. 
It vanishes in the limit of 
exact local supersymmetry transformations as it should.
%%%%%%%%%%%%%%%%%%%%%%%%%%%5
Terms in the second line of Eq.(\ref{matf}) can only be non-zero 
if the v.e.v. of the K\"ahler metric mixes 
fields from the visible sector with fields from the hidden sector. 
We shall not consider  this possibility here.  
%%%%%%%%%%%%%%%%%%%%%%%%%

\indent {\it ii)} terms that are proportional to the gravitino mass
(the last two lines of Eq.(\ref{matf})) and exist only 
in the framework of supergravity. They depend only on the structure 
of the K\"ahler potential. Of these terms the second
gives rise to a relatively suppressed 
Dirac neutrino mass and was used (in a different context) 
in Ref.\cite{Arnowitt}. 
Actually it is obvious that 
{\it all} terms in the third line of Eq.(\ref{matf}) can 
contribute to Dirac neutrino masses. The terms in the fourth line
of Eq.(\ref{matf}) require mixed hidden and  observable sector
kinetic terms and as with the terms in the second 
line of  Eq.(\ref{matf}) we assume they are absent.
%%%%%%%%%%%%%%%%%%%%%%%%%%%%%%%
%whereas the second term generates for example bilinear
%terms in the superpotential with magnitude
%$O(m_{3/2})$.
% It was this term that was 
%used in Ref.\cite{GM} to generate the $\mu$-term 
%
%Terms with a prefactor $1/\Mp^2$ in the third line in
%Eq.(\ref{matf}),
% are in general suppressed by two 
% ADDED TWO ABOVE 
%powers of the Planck
%scale and are thus negligible. 
They are only relevant  when $K^{(0)}$
and/or $W^{(0)}$ contain a tadpole gauge singlet.
% as in terms of the second line of Eq.(\ref{matf}). 

%SOME CHANGES ABOVE AND BELOW

%The fermion fields $\chi^\alpha$ in
%Eq.(\ref{fmatter}) need not be in the canonical basis. 
%However in cases where higher order operators in the K\"ahler potential
%have been derived (in 
%for example string theory in Ref.\cite{Antoniadis}) the various symmetries of 
%the theory dictate that their coefficients are
%of order one in the canonical basis and so 
%we henceforth confine our discussion to
%that case, $g^{ij*}=g^{\alpha\beta*}=1$.
% Although one can obtain enhanced neutrino masses by arranging 
%the K\"ahler metric to be
%close to singular, we shall not need to make use of such a peculiar
%mechanism in what follows.

%\setcounter{equation}{0}
%\section{Neutrino masses in Supergravity}
%\label{sec:sec3}

\section{(Pseudo-) Dirac  neutrino masses}

An obvious starting point for a theory of small Dirac neutrino masses
is to prevent them from appearing directly in the superpotential. A
natural solution to the $\mu$-problem~\cite{GM} would require in
addition the non-existence of the operator $H_u H_d$ in the
superpotential. This can naturally be done with a discrete
$R$-symmetry or perhaps
%%%$S$, 
some other symmetry. As a working example, let us consider an
$R$-symmetry with $R$-characters for the matter superfields given by 
Table~1. To
these we have added a right handed gauge singlet superfield $\bar{N}$
with $R$-character $R(\bar{N})=n$. The symmetry has to be chosen so
that the operators $L H_u \bar{N}+ H_u H_d$ are forbidden in the
superpotential but are present in the K\"ahler potential.
{In addition we will for definiteness suppose that the singlet $S$
has a non-zero $R$ character as well, 
so that its appearance in the superpotential will 
be limited as we'll see shortly. (Zero $R$-character for this singlet is also 
possible but {necessitates} other hidden sector fields.)}  
%%%%%%%%%%%%%%%%%%%%%%%%%%%%
\begin{table}[t]\begin{center}
\begin{tabular}{|c|c|c|c|c|c|c|c|}
\hline
$Q$ & $\bar{U}$ & $\bar{D}$ & $L$ & $\bar{E}$ & $\bar{N}$ & $H_u$ & $H_d$
 \\ \hline
$2-d-h$ & $d+2h$ & $d$ & $h-n$ & $2-2h+n$ & $n$ & $-h$ & $h
$\\ \hline
\end{tabular}
\caption{$R$-characters for the MSSM fields under the requirement
that the operators $H_u H_d+ L H_u \bar{N}+ L H_d^* \bar{N}$ appear only in the K\"ahler 
potential.}
\end{center}
\end{table}
%%%%%%%%%%%%%%%%%%%%%%%%%%%%%%%%%%%%%%%
%%%The hidden sector fields, $\sigma=S/M$ are $R$-neutral, $R(S)=0$. 
The  visible superpotential has $R(W)=2$. 
The    K\"ahler potential is $R$-neutral $R(K)=0$. 
We shall choose $n=-1$. For the moment we shall 
also assume lepton number conservation. The allowed terms  
are then 
%the usual MSSM trilinear  superpotential terms 
%$W^{\rm (o)}(\sigma,y)$ giving 
%masses to quarks and charged leptons, 
%and some additional terms in the 
%K\"ahler potential;
%%%%%%%%%%%%%%%%%%%%%%%%%%%%%
\begin{widetext}
%%%%%%%%%%%%%%%%%%%
\begin{eqnarray}
W^{\rm (o)}(\sigma,y) 
\ &\supset& \   
Y_E L H_d \bar{E}+ Y_D Q H_d \bar{D} 
+ Y_U Q H_u \bar{U} + W^\sigma
%+ \frac{g_4(\sigma)}{M}(L H_u) (L H_u)  
\;, 
%\nonumber \\ 
\label{sp} \\[2mm]
K^{\rm (o)}(\sigma,\sigma^*,y,y^\dagger) \ &\supset& \  
 c_1(\sigma,\sigma^*) H_u H_d  %\nonumber \\ &+&
+\frac{c_2(\sigma,\sigma^*)}{M} L H_u \bar{N} 
+ \frac{c_3(\sigma,\sigma^*)}{M}L H_d^* \bar{N}
+ {\rm H.c} \;,
\label{kp}
\end{eqnarray}
%%%%%%%%%%%%%%%%%%%%% 
%%%%%%%%%%%%%%%%%%%%%%%%%%%%%
\end{widetext}
%%%%%%%%%%%%%%%%%%%%%%%%%%%%
where $M$ is our ultraviolet cutoff {and $W^\sigma$ 
is the $\sigma$ dependent part of the superpotential which will be 
responsible for supersymmetry breaking (to be discussed later). As an example if $R(s)=2$ 
then this could be a Polonyi-like term $\beta S$ where 
$\beta$ is some constant. The
$c(\sigma,\sigma^*)$ coefficients are the result 
of all perturbative and non-perturbative contributions to the 
K\"ahler potential so we do not need to insist that $\sigma < 1$ 
although this is where we need to be to have perturbative control.
We may quite reasonably assume {these coefficients and their derivatives}
to be of order one}. Of course 
$W^{\rm (o)}$ and $K^{\rm
(o)}$ contain other
non-renormalizable terms, irrelevant to neutrino masses, of 
order $1/M$ and higher~\footnote{
If lepton number is violated, 
the $R$-character of the superfield $\bar{N}$ classifies the
additional neutrino mass operators and, although we do not present 
them here, models with 
other less phenomenologically appealing choices are possible.}. 
%Note that $R$-parity
%symmetry is preserved by the $R$-symmetry of Table~1 with $n=-1$. 

We can now use the master formula of  Eq.(\ref{matf}) 
together with Eq.(\ref{kp}) to obtain the relevant terms for the 
Dirac neutrino masses. Consider for simplicity one singlet, 
$\sigma $, and one generation of neutrinos with 
$\chi^\alpha = \overline{\nu_R}, 
\chi^\beta = \nu_L$\:;
%%%%%%%%%%%%%%%%%%%%%%%%%%%%%
%\begin{widetext}
%%%%%%%%%%%%%%%%%%%%%%%
%\begin{eqnarray}
%m_\nu^{\rm D} \ = \ \frac{m_{3/2}}{2}\: \Biggl [ \:
% \frac{\del^2 K^{\rm (o)}}{\del y^\alpha \del y^\beta} 
%&-& g^{\gamma\delta*}\frac{\del^3 K^{\rm (o)}}{\del y^\alpha\del y^\beta 
%\del y^{\delta*}}
%\frac{\del K^{\rm (o)}}{\del y^\gamma} 
%\nonumber \\[2mm] &-&  
%-\bigg(\frac{\Mp^2}{\langle W^{\rm (h)} \rangle 
%M^2} \biggr )
%\frac{\del^3 K^{\rm (o)}}{\del y^\alpha \del y^\beta \del
%\sigma^{i*}}\frac{\del W^{\rm (h)}}{\del \sigma^i} \: \Biggr ] \;.
%\label{matdf}
%\end{eqnarray}
%%%%%%%%%%%%%%%%%%%%%%%%%%%%%%%%%%%%
%%%%%%%%%%%%%%%%%%%%%%%%%%%%%
%\end{widetext}
%%%%%%%%%%%%%%%%%%%%%%%%%%%%%%
%Consider for simplicity one singlet, $\sigma $. 
%Inserting the higher order operators,
%the contributions from the terms in Eq.(\ref{matdf}) are found to be
%%%%%%%%%%%%%%%%%%%%%%%%%%%%%%%%
%\begin{eqnarray}
%m_\nu^{\rm D} \ &=& \ 
%v \: \biggl (
%\frac{m_{3/2} }{M}
%\biggr )
%\biggl [  \sin\beta \: c_2(\sigma,\sigma^*) - 
% \sin\beta \: c_1(\sigma,\sigma^*) c_3(\sigma,\sigma^*)
%\biggr ] 
%\nonumber \\[2mm] &-&  
%v \: \biggl (
%\frac{m_{3/2} }{M}
%\biggr )
%\biggl (\frac{\Mp}{M}\biggr )^2 
%\biggl ( \frac{1}{\langle W^{\rm (h)} \rangle}\frac{\del W^{\rm (h)}}{\del \sigma} \biggr ) \biggl 
%[ \sin\beta \frac{\del c_2(\sigma,\sigma^*)}{\del \sigma} 
%+\cos\beta \frac{\del c_3(\sigma,\sigma^*)}{\del \sigma} \biggr ] \;.
%\label{md}
%\end{eqnarray}
%%%%%%%%%%%%%%%%%%%%%%%%%%%%%
\begin{widetext}
%%%%%%%%%%%%%%%%%%%%%%%%%%%%%%%%%%%%%%%%%%%%%%%%%
%\begin{eqnarray}
%m_\nu^{\rm D} \ &=& \ 
%v \: \biggl (
%\frac{m_{3/2} }{M}
%\biggr )
%\biggl [  \sin\beta \: c_2(\sigma,\sigma^*) - 
% \sin\beta \: c_1(\sigma,\sigma^*) c_3(\sigma,\sigma^*)
%\biggr ] 
%\nonumber \\[2mm] &-&  
%v \: \biggl (
%\frac{m_{3/2} }{M}
%\biggr )
%\biggl (\frac{\Mp}{M}\biggr )^2 
%\biggl \langle \frac{ \del_\sigma W^{\rm (h)}}{ W^{\rm (h)} }
%\biggr \rangle 
%\biggl [ 
%\sin\beta \: \del_{\sigma^{\mbox{\tiny *}}}  c_2(\sigma,\sigma^*) 
%+\cos\beta \: \del_{\sigma^{\mbox{\tiny *}}}  
%c_3(\sigma,\sigma^*) \biggr ] \;,
%\label{md}
%\end{eqnarray}
%%%%%%%%%%%%%%%%%%%%%%Sakis%%%%%%%%%%%%%%%%%%%%%%%%%%%
\begin{eqnarray}
m_\nu^{\rm D} \ &=& \ 
v \: \biggl (
\frac{m_{3/2} }{M}
\biggr )
\sin\beta \: \biggl [   c_2(\sigma,\sigma^*) - 
  c_1(\sigma,\sigma^*) c_3(\sigma,\sigma^*)
\biggr ] 
\nonumber \\[2mm] &-&  
 v \: \biggl  (
\frac{ F_S}{M^2}
\biggr )
\sin\beta \:\biggl [ 
\del_{\sigma^{\mbox{\tiny *}}}  c_2(\sigma,\sigma^*) 
+\cot\beta \: \del_{\sigma^{\mbox{\tiny *}}}  c_3(\sigma,\sigma^*) \biggr ] \;,
\label{md}
\end{eqnarray}
%%%%%%%%%%%%%%%%%%%%%%%%%%%%%
\end{widetext}
%%%%%%%%%%%%%%%%%%%%%
where $\del_\sigma \equiv \del/\del\sigma $ and 
\begin{eqnarray}F_S=
\del_S W^{\rm (h)}+m_{3/2}\: \del_S K^{\rm (h)}\; .
\label{fs}
\end{eqnarray}
Notice that in
Eq.(\ref{md}) there exists a source for Dirac neutrino masses 
which survives even in the global supersymmetric limit. 
In this limit, only the term proportional to $\del_S W^{\rm (h)} $ 
remains and for 
$m_\nu^{\rm D} \ = \ (0.04 - 0.05)~{\rm eV}$ we find
%%%%%%%%%%%%%%%%%%%%
\begin{eqnarray}
\frac{F_{S}}{M^2} \ \simeq 
\frac{m^{\rm D}_\nu}{v \sin\beta \: ( 1 + \cot\beta)} = 
(1.6 - 2.8) \times 10^{-13} \;, \label{fss}
\end{eqnarray}
%%%%%%%%%%%%%%%%%%%% 
where we used $v=174.1$ GeV and $\tan\beta=1-60$ and have assumed that
the v.e.v. of all the $c$'s and their derivatives are unity (there is
the possibility of cancellation).  This is a rather model independent
result.  In local supersymmetry for example, vanishing of the
vacuum energy implies that
% we have 
$F_S=\sqrt{3} \Mp m_{3/2}$, and varying $100~{\rm GeV} <m_{3/2} <
10~{\rm TeV}$ we obtain
%%%%%%%%%%%%%%%
\begin{eqnarray}
4\times 10^{16}~{\rm GeV} < M < 5 \times 10^{17}~{\rm GeV}\, .
\label{eq11}
\end{eqnarray} 
%%%%%%%%%%%%%%%
The terms in the second line of Eq.(\ref{md})
are enhanced by a factor $\Mp/M$ 
relative to the terms in the
first line, and thus are the dominant ones for any scenario.
This is important for example in no-scale models where 
the gravitino can be quite light. In addition, one should note
that the new non-holomorphic term proportional to $\partial_{\sigma^*}c_3$
dominates in Eq.(\ref{md}) if $c_1$ and $c_2$ take on small values.

{We should remark that the soft breaking masses of other particles
such as squarks are proportional to $\tilde{m} \sim F_S/M$ 
 making  them generically somewhat larger than desirable, which requires
some degree of tuning.} For example if there are terms
$c_i(\sigma,\sigma^*)\, \phi_i^\star \phi_i$ in the K\"ahler potential
then for $M=M_{\rm GUT}$ we obtain $\tilde{m}_i \sim 3~{\rm
TeV}-6~{\rm TeV}$ and for $M=M_{\rm s}$ we obtain $\tilde{m}_i \sim
100$ TeV.  A suppression of the $\sigma$ dependence in the operator
$\phi_i^*\phi_i $ in the K\"ahler potential is therefore required. For
example it could be forbidden by hidden sector symmetries.
% This is clearly more serious for $M\sim M_{\rm{s}}$ than for
% $M\sim M_{GUT}$ where a suppression of maybe a factor of 3 is
%necessary.  However 
This is model dependent and we will not present a
detailed discussion of this as we would like to preserve our
phenomenological approach.

% the role of the non-holomorphic term in the 
%K\"ahler potential proportional to $c_3(\sigma,\sigma^*)$ .
%Putting in numbers, ($m_{3/2}=1$ TeV, $M
%\simeq 3 \times 10^{17}$ GeV, $v=174.1$ GeV, $\tan\beta=10$) and
%assuming that all the $c's$ and their derivatives are unity (bare in
%mind cancellations), we obtain
%%%%%%%%%%%%%%%%%%%%%%%%%%%%%%%%%%%%%%%%%%%%%%%%%%
%\begin{eqnarray}
%m_\nu^{\rm D} \ &=& \ 0.04~{\rm eV} \;.
%\label{nmass}
%\end{eqnarray}
%%%%%%%%%%%%%%%%%%%%%%%%%%%%%%%%%%%%

We emphasize that the constraint of                                           
Eq.(\ref{fss})  survives in the global supersymmetric limit.                           
This is an important condition for incorporating  Dirac neutrinos  
in models  with low scale supersymmetry breaking  as               
for example in the case of gauge mediated supersymmetry breaking.

The values obtained for $M$ in Eq.(\ref{eq11}), naturally lie between $M_{GUT}$ and 
the heterotic string scale $\Ms$ for a very wide range of parameters,
the result anticipated in the introduction. 
Note that the neutrino mass in Eq.(\ref{md}) varies as the
square of $M$ so that the value is rather accurately determined.
This is the central point of this paper, that in
supergravity small neutrino masses of the experimentally
observed size can arise, 
%automatically,
and that neutrinos are predominantly
Dirac fermions.  In contrast, the operation of a see-saw mechanism
demands the introduction of extra scale(s) in order to obtain  the 
correct order of magnitude. Of course, the input in our
case was an $R$-symmetry which forbade direct neutrino masses.
However this requirement is more general than the neutrino mass
problem at hand, since it is also necessary to resolve the $\mu$-problem. 

%Reversing the argument,
%if we assume that the atmospheric neutrino masses are
%given by
%Eq.(\ref{md}), take into acount the experimental fact that
%$m_\nu=0.04-0.05$~eV, and allow the gravitino mass to be in the 
%range $500~{\rm
%GeV}<m_{3/2}<3~{\rm TeV}$, we then find that the scale $M$ has to be
%%%%%%%%%%%%%%%%%%%%%%%%%%
%\begin{eqnarray}
%2.4\times 10^{17}~{\rm GeV} < M < 4.4 \times 10^{17}~{\rm GeV} \;
%\end{eqnarray}
%%%%%%%%%%%%%%%%%%%%%%%%5which is suspiciously close to the heterotic string scale as we have
%which remarkably is the heterotic string scale, 

%\subsection{Pseudo-Dirac Neutrino masses}

If we relax the assumption of lepton number conservation
then the Dirac neutrinos obtained from the K\"ahler potential can be
``polluted'' by the presence of active Majorana neutrino 
masses derived from extra non-renormalizable terms in 
addition to those in  Eqs.(\ref{sp},\ref{kp});
% and to $W^{\rm (o)}(\sigma,y)$ of the MSSM  

\begin{eqnarray}
W^{\rm (o)}(\sigma,y) 
\ &\supset& \   
%Y_E(\sigma) L H_d \bar{E}+ Y_D(\sigma) Q H_d \bar{D} 
%+ Y_U(\sigma) Q H_u \bar{U} 
 \frac{g_4(\sigma)}{M}(L H_u) (L H_u)  
\;, 
%\nonumber \\ 
\label{sp2} \\[2mm]
K^{\rm (o)}(\sigma,\sigma^*,y,y^\dagger) \ &\supset& \  
% c_1(\sigma,\sigma^*) H_u H_d  %\nonumber \\ &+&
%+\frac{c_2(\sigma,\sigma^*)}{M} L H_u \bar{N} 
%+ \frac{c_3(\sigma,\sigma^*)}{M}L H_d^* \bar{N}
\frac{c_4(\sigma,\sigma^*)}{M^3}W^{(h)}\bar{N}^2
+ {\rm H.c.} \;
\label{kp2}
\end{eqnarray}
Assume that only the first term is
present. (The second term is 
quite high order to get a zero $R$-charge, so one could argue that 
such terms become suppressed.) 
%~\footnote{The last term in Eq.(\ref{sp}) can be prohibited 
%by simply assuming lepton number conservation.},
%A lepton number discrete symmetry can be imposed,
%$L_3=e^{2\pi a_3/3}$, $a_3=0,0,0,-2/3,2/3,2/3,0,0$ to the superfields
%in Table~1 respectively so that Majorana masses do not arise at all.
Then from the first term in Eq.(\ref{matf}) with $\chi^\alpha
= \overline{\nu_L^c},
\chi^\beta = \nu_L$ we obtain 
%%%%%%%%%%%%%%%%%%%%%%%%%%
\begin{eqnarray}
m_\nu^{\rm L} \ =\ g_4(\sigma)\:  
\frac{ v^2 }{ M} \: \sin^2\beta\;.
\label{mm}
\end{eqnarray}
%%%%%%%%%%%%%%%%%%%
%Assuming as an  example a Polonyi type superpotential 
%$W^{\rm (h)}=\Mp^2 m_{3/2} (\sigma +a)$ and the parameters used in 
%the previous paragraph we obtain, 
%%%%%%%%%%%%%%%%%%%%%%%%%%
%\begin{eqnarray}
%m_\nu^{\rm L} \ =\ g_4 v (\frac{v}{M}) \sin^2\beta \;.
%\end{eqnarray}
%%%%%%%%%%%%%%%%%%%%%%%
%From the definition of $m_{3/2}$ the resulting mass 
%is actually independent of 
%$W^{(h)}$ as long as $\langle K \rangle < \Mp^2 $.
For the range of $M$ above 
we obtain $m_\nu^{\rm L} = (3\times 10^{-5}-7\times 10^{-4})$ eV.
In summary, Dirac and Majorana neutrino masses
(we consider one generation of neutrinos)  are combined in
the basis $(\chi = \nu_L+\nu_L^c, \omega = \nu_R+\nu_R^c)$
%%%%%%%%%%%%%%%%%%%%%%%%%%%
\begin{eqnarray}
\label{pseudomat}
(\overline{\chi}\,\,\,\, \overline{\omega}) \,
\biggl ( \begin{array}{cc} m_\nu^{\rm L} & m_\nu^{\rm D} \\
                           m_\nu^{\rm D}             & 0 
\end{array} \biggr ) 
\biggl ( \begin{array}{c} \chi \\
                          \omega 
         \end{array}
\biggr )
\;, \label{eq217}
\end{eqnarray}
%%%%%%%%%%%%%%%%%%%%%%% 
with eigenvalues close to $m_\nu^{\rm D}$. The small mass 
splitting  between the two physical eigenstates is
%%%%%%%%%%%%%%%%%%%%%%%%%
\begin{eqnarray}
\delta m^2 \simeq 2 m_\nu^{\rm D} m_\nu^{\rm L} &=& (3\times 10^{-6}
- 5 \times 
10^{-5}) ~{\rm eV}^2 \; ,
\end{eqnarray}
%%%%%%%%%%%%%%%%%%%%%%%%%
and the mixing angle $\tan 2 \theta \ =\ 2 \frac{m_\nu^{\rm
D}}{m_\nu^{\rm L}} $ very close to maximal, $\sin 2 \theta = 1$.  
Thus, neutrinos are pseudo-Dirac~\cite{pd,Lim}; the Dirac neutrino splits
into a pair of two {\it maximally} mixed Majorana neutrinos with 
almost equal masses. Furthermore, the effective mass for the neutrinoless
double beta decay is given by~\cite{Bell},
%%%%%%%%%%%%%%%%%%%%%%%%%
\begin{eqnarray}
\langle m_{\rm eff}\rangle 
 \ = \  \frac{1}{2} \sum_j U_{ej}^2 
\frac{\delta m^2_j}{2 m_j}\;, \label{0bbnu}
\end{eqnarray}
%%%%%%%%%%%%%%%%%%%%%%%%%
where $U$ is the neutrino mixing matrix determined by the solar
and atmospheric neutrino oscillations. Using the numbers quoted above, 
we find that 
Eq.(\ref{0bbnu}) 
gives $\langle m_{\rm eff} \rangle = 
(10^{-5} - 3 \times 10^{-4}) $ eV. 
One cannot detect neutrinoless double $\beta $ decay of such small magnitudes, 
and these contributions are therefore unobservable for the forseeable future. 
One may instead have to resort to astrophysical techniques to distinguish 
pseudo-Dirac from Dirac neutrinos~\cite{Bell}.  
%We should remark in passing that is difficult to distinguish between the two 
%Majorana particles of which the Pseudo-Dirac neutrino is composed. 
%% EH ? ISN'T ONE STERILE AND THE OTHER ACTIVE?
Furthermore, if the $\bar{N}^2 $ operator of Eq.(\ref{kp2}) 
is present and of equal size, 
%the 
%Majorana neutrino masses $m_\nu^{\rm R}$ 
%generated will fill out the 22 element of the 
%mass matrix. The terms are of the same order as the Dirac terms, 
the situation becomes highly involved, 
with the three generations of neutrinos having a general $6\times 6$ 
mass matrix. 
%Phenomenological and astrophysical problems and probes  of pseudo-Dirac
%neutrinos have been discussed recently in Ref.\cite{Lim,Bell}.

%The inputs used in this section should be considered as indicative. 
%Indeed, there is no good motivation (today) for a Polonyi
%potential ansatz. Indeed, with another ansatz (maybe motivated from string
%theory ) one could obtain much smaller mass splitting, $\delta m^2$ for
%the pseudo-Dirac pair. 
%However, the pattern of the matrix Eq.(\ref{eq217})
%with $m_\nu^{\rm D}>>m_\nu^{\rm L}$ is rather general in the supergravity 
%scenario with naturally light (Pseudo)Dirac neutrino spectrum explained above. 

\section{Questions and Conclusions}

There are a number of questions that arise. The most pressing concerns the 
source of the non-renormalizable terms in the K\"ahler potential of 
Eq.(\ref{kp}).
%It is known that non-renormalizable neutrino mass terms 
%can be generated in various ways even in global supersymmetry~\cite{Brignole}. 
%However this makes no use of the fact that $M\simeq \Ms $ which we find rather 
%appealing. 
The analysis presented here leads us to suspect that the required
operators may appear simply as effective operators in heterotic string
theories in much the same way as the $\mu$ term
does~\cite{Antoniadis}.  {The scale $M$ may also appear
radiatively in the K\"ahler potential, along the lines discussed
in~\cite{Brignole} or explicitly by construction in a GUT model.}  One
aspect of this picture that we find appealing is that, in contrast
with the see-saw picture, the connection with string or GUT scale
physics is rather immediate. The neutrino masses and mixings are not
filtered through unknown Majorana terms but carry direct information
about the structure of the K\"ahler metric. {This fact  certainly offers 
new opportunities for neutrino model building.}   

{ In this paper we have been arguing that the scale of neutrino
masses may quite easily be associated with the scale of supersymmetry
breaking and hence the Weak/Planck scale hierarchy, in the very same
way that the $\mu$-term can. Although this is a general observation,
we are obliged to present a simple model of supersymmetry breaking
where $F_S$ is generated with the correct size with the charges we
have been using.  Consider for example an $R$-charge for the singlet
$R(S)=1$. In this case the supersymmetry breaking part of the
potential can take the form
\[
W^\sigma = \beta S^2 
\]
where $\beta$ is a dimensionful coupling of order $M_W$. The fact that
this represents a fine-tuning is of course the {\em usual} tuning
problem associated with supersymmetry breaking. $S$ needs to get a
v.e.v and in order for this to happen we may further suppose that the
$R$-symmetry we are using is gauged and anomalous. Such models were
considered in ref.\cite{Dreiner}, and it is known that such a symmetry
must be broken at scales $M \le \Mp$, and that there are no effects
from gauging the $R$-symmetry remaining at low energies. Because of
the $R$-charge of $S$ it is now perfectly natural for $S$ to get a
v.e.v of order $M$ from the Fayet-Iliopoulos $D$-term of the
$R$-symmetry, especially as it has no other $D$ terms to force it to
zero v.e.v.  This then gives
\[
\langle 
W^\sigma \rangle \sim \beta M^2 \,\,\, ; \,\,\, F_S \sim \beta M^2 \frac{K_S}{\Mp^2} + W_S \sim \beta M\, .
\]
The value of $F_S$ may now be tuned to $\sqrt{3} M_W \Mp$ get zero cosmological constant 
as usual. But the point 
is of course that we now have to make {\em no additional 
tuning to get the Dirac neutrino masses of the right order} and this
is the main finding of the paper.}

{Another important
question is how to account for non-trivial (maximal) neutrino mixing
matrix.  The answer to this question may be linked to the fact that
the K\"ahler potential parameters are not protected by the
non-renormalization theorem, and vertex corrections may induce large
flavour mixing through Renormalization Group running. }

To summarize, we have shown that minimal supergravity naturally allows 
Dirac masses without the ad-hoc addition of any new mass scales. 
If there is lepton number conservation, then 
the MSSM naturally contains pure Dirac neutrino masses 
that are comparable to the atmospheric neutrino mass. 
The only other remnant would be a slowly 
decaying right-handed s-neutrino with mass $\sim $ 1 TeV. 
%Regarding flavour and mixing,
%(which we haven't touched upon at all)
%it is clear that in this case 
%it should be much closer to the usual ``CKM-type'' phenomenology,  
%than see-saw neutrinos. 
We have throughout been focussing on the 
atmospheric neutrino mass, but the remaining 
masses and mixings could be generated by the Yukawa couplings in 
the K\"ahler potential of Eq.(\ref{kp}) in much the same way 
as the quark masses and mixings.    
If lepton number is violated, then we have seen that it is possible to 
get either pseudo-Dirac neutrinos or a general $6\times 6$ Majorana 
mass matrix structure with naturally small elements.
Finally, we should remark that baryogenesis can be 
accommodated via leptogenesis with Dirac neutrinos~\cite{Lindner}. 
%Actually, the extensions required for this are minimal and it may even be 
%possible to adapt the idea for the MSSM without modification.

% For the moment however, minimal supergravity  
%phenomenology with the full 6x6 neutrino 
%mass matrix  remains to be explored. 
%Spontaneously broken supergravity induces the neutrino
%mass operators of Eqs.(\ref{sp},\ref{kp}).
%These neutrino oporators  are renormalized
%from the scale $M$ all the way down to low energies. 
%Neutrino masses are generated after the (radiative) 
%electroweak symmetry breaking, in  Eq.(\ref{md}). 
%In addition, broken local supersymmetry induces 
%soft breaking masses and couplings. .........

%In this Supergravity scenario,
%right handend s-neutrinos have masses of the order of
%$100~{\rm GeV} \div 1~{\rm TeV}$ and decay only 
%gravitationally to the the lightest supersymmetric particle 
%which we assume is the neutralino. 

%What is finally the model : soft L, running
%some pheno with psedodirac

%Some Cosmo with the sterile sneutrinos

%a natural model

{\it  Acknowledgements} - We would like to thank Ignaccio Navarro 
for very useful comments on the 
manuscript and Sacha Davidson and Subir Sarkar for useful discussions. 
KT would like to thank the IPPP for 
hospitality during this work.

%%%%%%%%%%%%%%%%%%%%%%%%%%%%%%%%%%%%%%%%%%%%%%%%%%

\end{document}